\documentclass[article]{emulateapj}

\usepackage{graphicx}
\usepackage{txfonts}
\usepackage{natbib}

\usepackage{verbatim}
\usepackage{longtable}
\usepackage{lscape}
\usepackage{afterpage}
\usepackage{multirow}
\usepackage{verbatim}

\newcommand{\neII}{[Ne\,\textsc{ii}] }

\newcommand{\hII}{H\,{\sc ii}~}

\begin{document}

\title{Mid-infrared PAH and H$_2$ emission as a probe of physical conditions in extreme PDRs}

\author{%
O. Bern\'e \altaffilmark{1}, 
A. Fuente \altaffilmark{2},
J. R. Goicoechea \altaffilmark{1},
P. Pilleri \altaffilmark{3,4},
M. Gonz\'alez-Garc\'ia \altaffilmark{5},
C. Joblin \altaffilmark{3,4}
}


\altaffiltext{1}{Centro de Astrobiolog\'ia (CSIC/INTA), Laboratiorio de Astrof\'isica Molecular, Ctra. de Torrej\'on a Ajalvir, km 4
28850, Torrej\'on de Ardoz, Madrid, Spain}

\altaffiltext{2}{Observatorio Astron\'omico Nacional, Apdo. Correos 112, 28803 Alcal\'a de Henares (Madrid), Spain} 

\altaffiltext{3}{Universit\'e de Toulouse ; UPS ; CESR ; 9 ave colonel Roche, F-31028 Toulouse cedex 9, France}

\altaffiltext{4}{CNRS; UMR 5187; 31028 Toulouse, France}

\altaffiltext{5}{ LUTH, Observatoire de Paris and Universit\'e Paris, 7 place Jansen, 92190 Meudon, France}

\begin{abstract}

Mid-infrared (IR) observations of polycyclic aromatic hydrocarbons (PAHs) and molecular hydrogen emission
are a potentially powerful tool to derive physical properties of dense environments
irradiated by intense UV fields. 
We present new, spatially resolved, \emph{Spitzer} mid-IR spectroscopy of the 
high UV-field and dense photodissocation region (PDR) around Monoceros R2, the closest ultracompact \hII
region, revealing the spatial structure of ionized gas, PAHs and H$_2$ emissions. Using a PDR model and 
PAH emission feature fitting algorithm, we build a comprehensive
 picture of the physical conditions prevailing in the region. 
 We show that the combination of the measurement of PAH ionization fraction and  
 of the ratio between the H$_2$ 0-0 S(3) and S(2) line intensities, respectively at 9.7 and 12.3 $\mu$m, 
 allows to derive the fundamental parameters driving the PDR: temperature, density and UV radiation field
 when they fall in the ranges $T = 250-1500\,$K, $n_H=10^4-10^6$cm$^{-3}$, $G_0=10^3-10^5$ respectively. 
These mid-IR spectral tracers thus provide a tool to probe the similar but unresolved UV-illuminated surface of protoplanetary disks
or the nuclei of starburst galaxies.

\end{abstract}


\section{Introduction} \label{int}

Dense ($>10^4$ cm$^{-3}$) and high UV field ($>10^4$
times the standard interstellar radiation field in units of the Habing field, written $G_0$
hereafter) photodissociation regions rule the energy balance, and thus evolution,
of some of the most fundamental astrophysical objects such as
protoplanetary disks, planetary nebulae and starburst galaxies.
One of the best tools to probe this UV illuminated matter is spectroscopy in the
mid-infrared (5-15 $\mu$m, hereafter mid-IR), because it provides information on both the gas
and small dust grain properties (polycyclic aromatic hydrocarbons and very 
small grains, hereafter PAHs and VSGs) . 
Ideally, one would like to spatially resolve the emission of these components
in order to study their variations as the UV field is attenuated.
Unfortunately, such observations are 
very hard to achieve (for the moment) on one hand because large ground-based 
telescopes, providing arcsecond angular resolution, are restrained in wavelength
coverage, while on the other hand, space borne telescopes have diameters that
are usually too small to resolve the sources.
Because it is the closest ultra compact \hII region at a distance of 850 pc,
Monoceros R2 (Mon R2, see \citealt{woo89} and \citealt{how94})
constitutes one of the rare exceptions where one can resolve the PDR
 between the \hII region and molecular cloud. In this Letter, we present and analyze 
 the mid-IR PAH and molecular hydrogen emissions in Mon R2 based on 
spatially resolved \emph{Spitzer} spectral mapping.

\begin{figure*}
\begin{center}
\epsscale{1.0}
\plotone{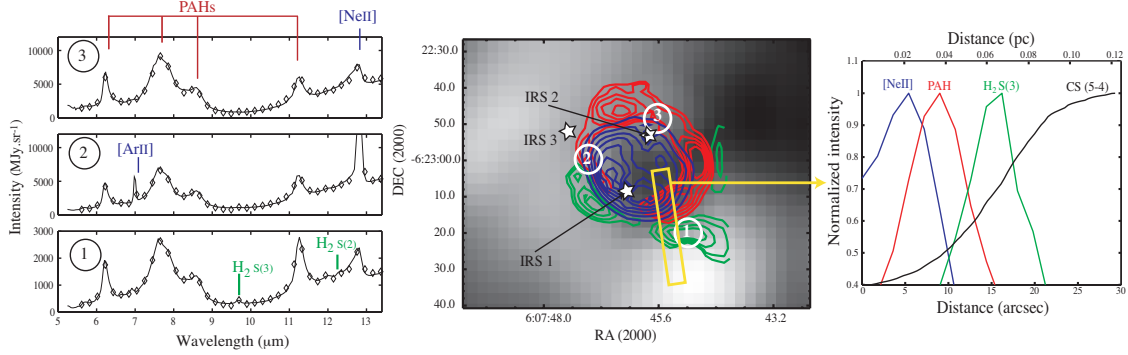}
\vspace{-0.0cm}
\caption{Overview of the Mon R2 region seen with \emph{Spitzer}. \emph{Left:}
Observed IRS spectra for PDR 1, 2 and 3 (continuous line) and fit by the model (diamonds, see Sect.\ref{mim}). 
\emph{Middle:}  The blue contours represent the intensity of the \neII line (0.02 to 0.1 erg s$^{-1}$ cm$^{-2}$sr$^{-1}$ in linear steps), the red contours the intensity of the  PAH 11.3 $\mu$m band (1.8 to 3.4 $\times 10^{-2}$ erg s$^{-1}$ cm$^{-2}$sr$^{-1}$ in linear steps) and the green 
contours the intensity of the H$_2$ 0-0 emission in the S(3) rotational line at 9.7 $\mu$m (1.5 to 4.5 $\times 10^{-4}$ erg s$^{-1}$ cm$^{-2}$sr$^{-1}$ in linear steps). 
The background map shows the CS $J$=5-4 emission presented in \citet{cho00}. Low emission is in black.  Numbers indicate the positions of the selected PDRs. \emph{Right:} Cut 
in the map along the yellow box, showing the stratified emission of the different lines.
\label{Mfit} }
\end{center}
\end{figure*}

\section{Observations} \label{obs}

Mon R2 was observed using the Infrared Spectrograph (IRS) onboard \emph{Spitzer}, in the low
resolution mode (${\lambda}/{\Delta~\lambda}=60-127$) as part of the "SPECHII" program (PI C. Joblin).
The data were obtained in the spectral mapping mode. The full spectral cubes of the SL1 and SL2 modules
were built using the CUBISM software \citep{smi06} from the basic calibrated files retrieved from the 
\emph{Spitzer} archive (version S19 pipeline). The two cubes were then assembled to provide 
the full SL cube of 26$\times$36 positions in space and $\sim$ 170 points in wavelength, 
ranging from 5 to 14.5 $\mu$m.

\section{Observational results} \label{ore}

An overview of our observations is presented in Fig.\ref{Mfit}.
The ionized gas in the \hII region, traced by the \neII line at 12.8 $\mu$m,  appears confined in a spherical region, as seen
by radio continuum observations \citep{woo89} and in higher angular and spectral resolution \neII ground based 
observations \citep{tak00, jaf03}. 
The cometary shape of the \hII region is well seen in our \neII map and peaks at about 0.12 erg  s$^{-1}$  cm$^{-2}$ sr$^{-1}$ 
in the surroundings of the B1 star IRS1. The fact that \neII maxium contours 
do not include the exact position of IRS1 is due to the saturation of the IRS detectors
at this position, implying that the intensity could not be measured, though likely peaking there
as shown by \citet{tak00} and \citet{jaf03}. PAH emission is present 
everywhere in the region, but for clarity in Fig.\ref{Mfit} we only present the region 
where it is the brightest in the 11.3 $\mu$m band (1 to 4 $\times 10^{-2}$ erg s$^{-1}$ cm$^{-2}$sr$^{-1}$)
forming a filamentary/shell structure that surrounds the \hII region. The  H$_2$ 0-0
S(3) rotational line at 9.7 $\mu$m peaks in a filament which lies between the \hII region
and the cold and dense molecular gas traced by CS $J$=5-4 (Fig.~\ref{Mfit}) transition. The 
intensity of the H$_2$ 0-0 S(3) line is of the order of 1-4 $\times 10^{-4}$ erg s$^{-1}$ cm$^{-2}$ sr$^{-1}$.
The H$_2$ 0-0 S(2) line at 12.3 $\mu$m follows a similar spatial distribution. As in the Orion bar \citep{tie93},
the spatial distribution of these H$_2$ lines is not correlated with the PAH emission contrary to what
is seen in lower UV field PDRs like the Horsehead nebula \citep{hab05, com07} or the $\rho$-Ophiucus filament \citep{hab03}.
In the following we investigate the origin of these structures from a physical point of view,
using PAH emission and modeling the H$_2$ excitation in the PDR.  To simplify the task we 
select three different zones named PDR1, PDR2, and PDR3 as
displayed in Fig.\ref{Mfit}. PDR1 was chosen to be representative of the region
where the maximum of H$_2$ emission is found (PAH emission is weaker but present). PDR2 is the transition from
the \hII region to the molecular gas. At these two positions it is well seen that H$_2$ gas
is found further from the star than the PAH emission (see cut in Fig.\ref{Mfit}).
Finally, we positioned PDR3 in a region situated further from IRS1 which is
a smoother transition from ionized to neutral and molecular gas.
For comparison, we also consider the mid-IR spectrum obtained integrating 
the whole IRS cube over the area imaged.

\section{Probing the physical conditions of a high UV field, high density PDR}

\subsection{H$_2$ emission as a probe of gas density and temperature}\label{test}

\begin{figure}
\begin{center}

\includegraphics[width=\hsize,angle=0]{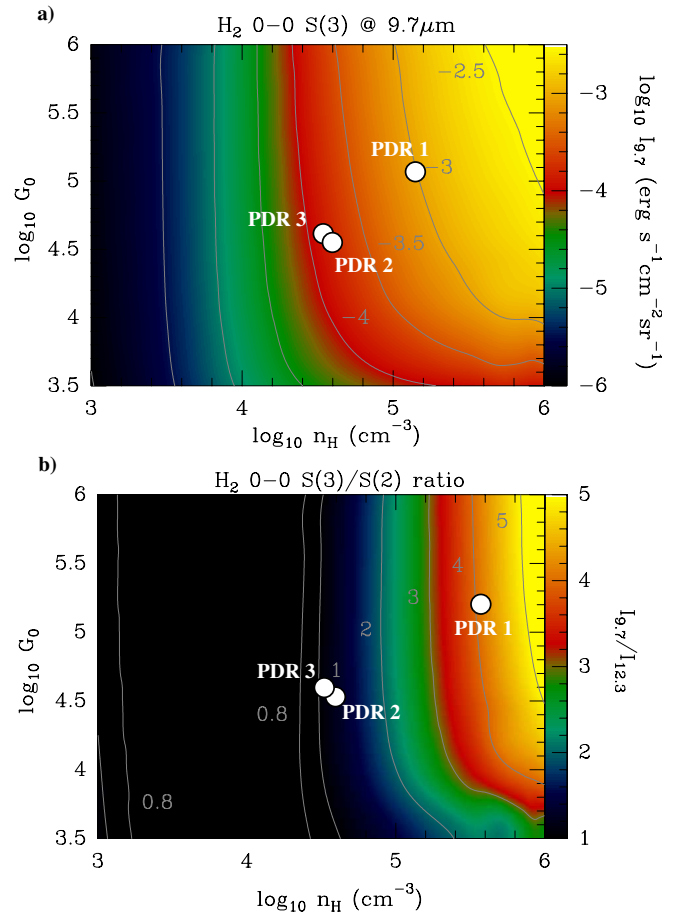}

\caption{ Results from the Meudon PDR code for high UV field/density conditions.
{\bf a)} Variations of the H2 0-0 S(3) absolute intensity (colors and contours) as a function of the hydrogen nuclei density $n_H$
and intensity of the radiation field $G_0$.
{\bf b)} Variations of the S(3)/S(2) line ratio (colors and contours) as a function of $n_H$
and $G_0$. Circles indicate the intersection between observed values of $I_{S(3)}$ (panel {\bf a)})
and $I_{S(3)}/I_{S(2)}$ (panel {\bf b)}) and the $G_0$ derived using PAH ionization fraction for PDR 1,2 and 3.
 \label{pdr}}
\end{center}
\end{figure}

In dense and high-UV PDRs like Mon R2, the excitation of the H$_2$ pure rotational lines
is dominated by inelastic collisions \citep{leb99}. To quantitatively probe the role of density $n_H$ 
and  radiation field $G_0$ on the excitation of the lowest energy rotational levels of H$_2$, we use the revised Meudon PDR code 
\citep{lep06, goi07}, for a large grid of radiation fields and densities above $10^3$ cm$^{-3}$.
Since the critical density of the S(3) line is $\sim 5\times10^4$ cm$^{-3}$  \citep{leb99}
the lowest energy rotational levels are thermalized when  $n_{H} \gtrsim 5\times10^4$ cm$^{-3}$.
As a consequence the S(3)/S(2) line ratio scales with the gas temperature
 and $T_{rot}\simeq T$, where $T_{rot}$ is the rotational temperature of the associated
 rotational transition. In high UV high density PDRs,
the main mechanism heating the gas is the photo-electric heating
by electrons ejected from very small dust grains and PAHs.
Thus, the photoelectric heating efficiency will depend on the ability to eject electrons
 from the  grain surface (less efficient as grains become positively
charged) and on the density of electrons in the gas with which charged grains recombine
and neutralize.  In PDRs, low energy electrons 
are provided by the ionization of carbon atoms, and as essentially all the carbon is
ionized, the electron density depends on gas density and carbon abundance. 
Overall, as the elemental abundace [C]/[H] is constant, an increase
of the gas density (thus of electron density) reduces the grains charge and increases
the photoelectric heating efficiency (i.e., the gas temperature).
This explains the dependence of the S(3)/S(2) ratio with density seen in Fig.\ref{pdr}. 

\begin{figure}
\begin{center}
\includegraphics[width=6cm]{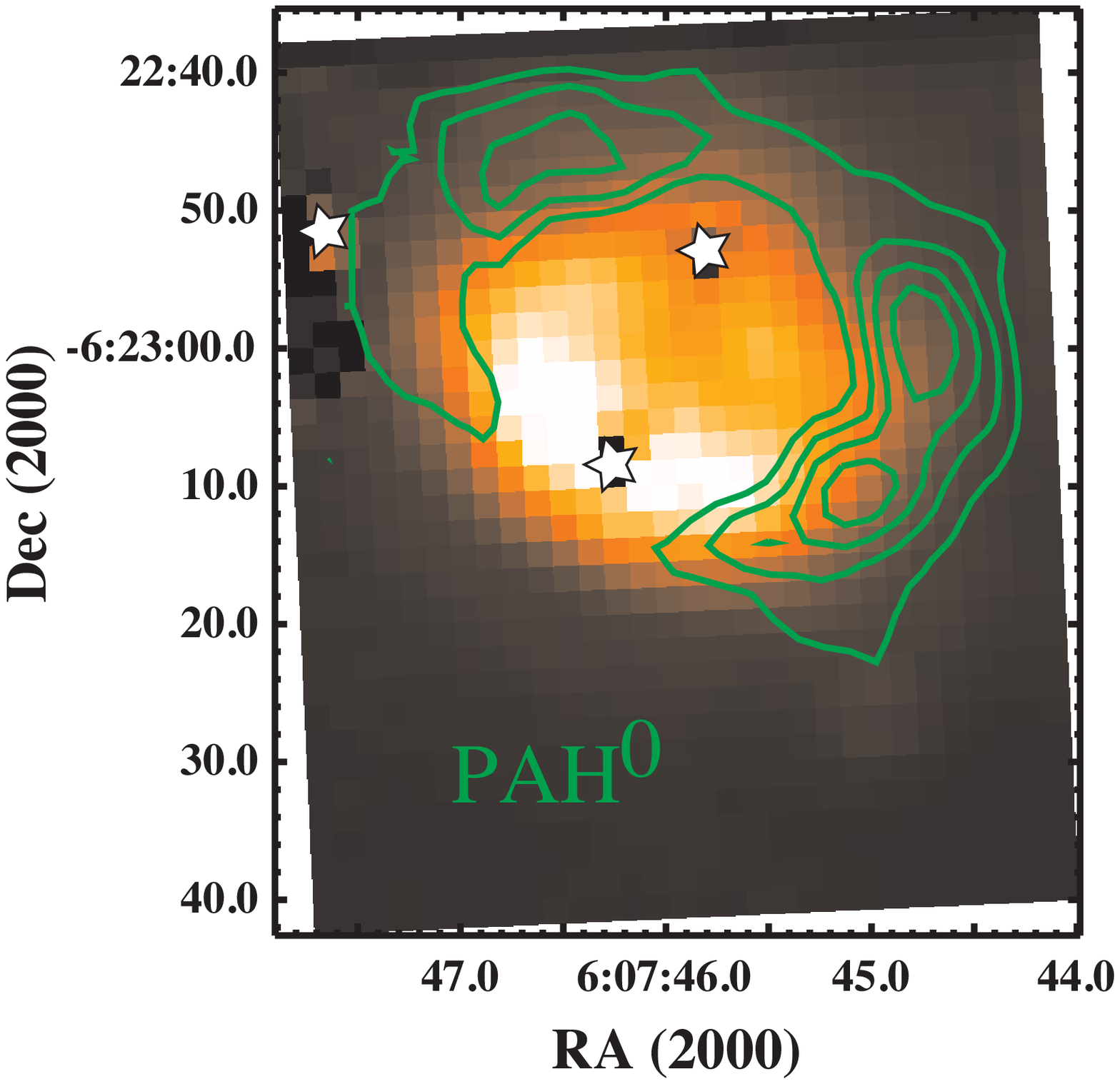}
\includegraphics[width=6cm]{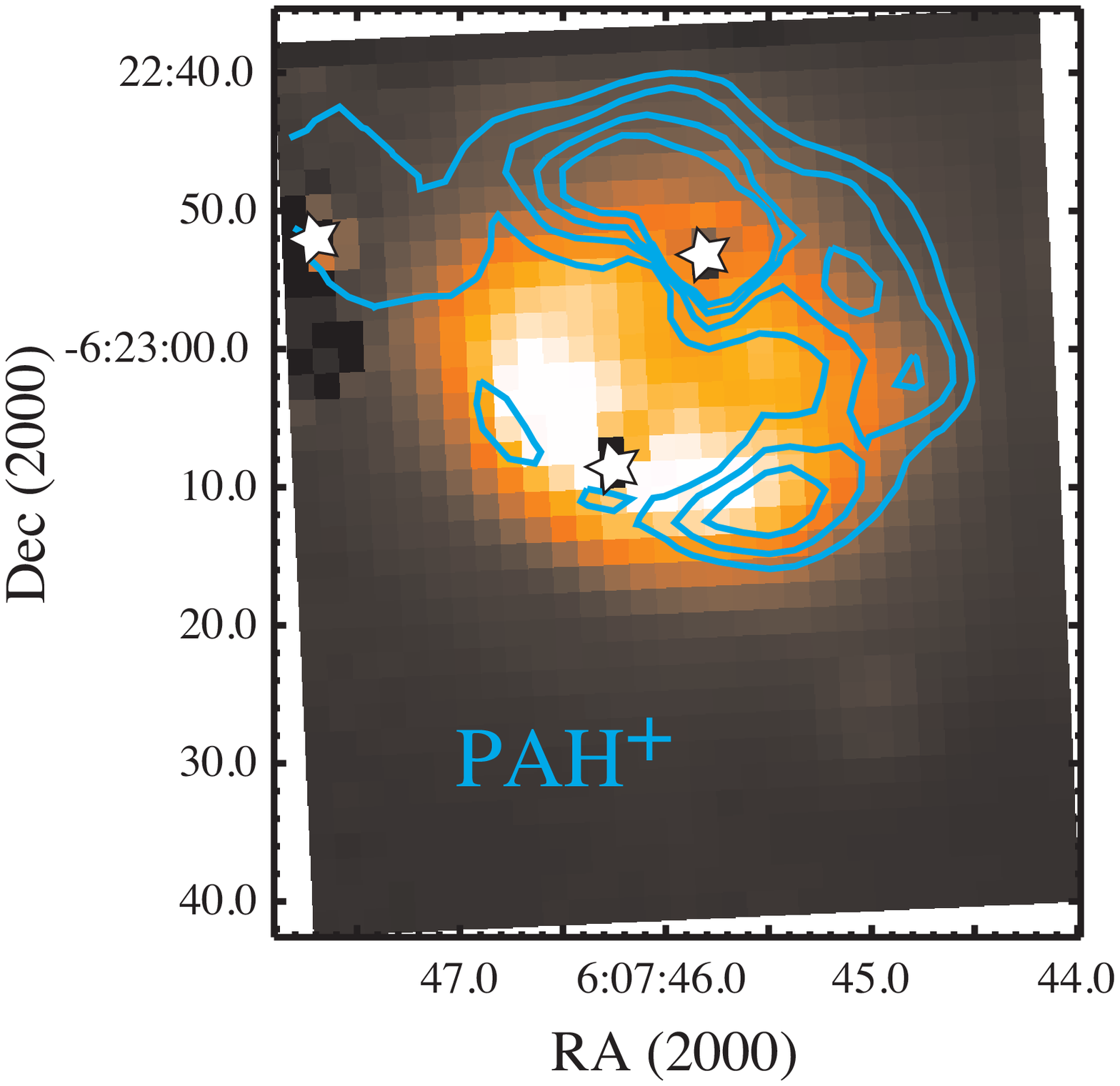}
\caption{ Contours of PAH$^{0}$ and PAH$^+$ intensity (resp. green and blue) overlayed on [Ne {\sc ii}] intensity map,
built from the results of the fit obtained for each spectrum of the cube of Mon R2. Positions of IRS 1,2,3 marked as in Fig.~\ref{Mfit}
\label{MonR2_map}}
\end{center}
\end{figure}

\subsection{PAHs as a probe of radiation field} \label{ove}

\subsubsection{Mid-IR emission model}\label{mim}

In order to analyse the full \emph{Spitzer} mid-IR cube we use a model
adapted from \citet{job08} and \citet{ber09}. This emission model includes the four major dust components,
 introduced in these previous works: VSGs (small carbonaceous dust grains), free neutral PAHs,
 positively ionized PAHs (PAH$^+$), and finally
PAH$^x$ that are large (with $N_C>100$  carbon atoms) charged PAHs. We decompose the observed mid-IR emission
using the 4 template spectra presented in \citet{ber09}. In addition, we consider continuum emission
that is simply modelled with two slopes. 
Furthermore, we include the effect of extinction by multiplying the mid-IR spectrum by
$({1-e^{-\tau_{\lambda} }})/{\tau_{\lambda}} $ where $\tau_\lambda = C_{ext}(\lambda) \cdot N_{l}$,
 $C_{ext}(\lambda)$ being the extinction cross section per nucleon, taken from \citet{wei01} with $R_V = 5.5$
 and $N_l$, the column density on material on the line of sight is left as a free parameter in the fit.
The  values A$_V$ of visual extinction corresponding to the derived $N_l$ (related by $N_l=1.8\times10^{21}A_{V}$)
for each PDR, and used to correct the measured H$_2$ lines intensities as a function of 
their wavelength, are given in Table \ref{table2}. We fit all the spectra of the cube using this technique,
thus providing the spatial distributions of the emission of each component. 
An example of these fits is provided in Fig.~\ref{Mfit} and spatial distributions of PAH$^+$ 
and PAH$^0$ populations are presented in Fig.~\ref{MonR2_map}.


\begin{table*}[ht!]
\caption{Observational diagnostics in PDR1, PDR2 and PDR3. Observational errors are given for H$_2$ lines.}
\label{table2}
\begin{center}
\begin{tabular}{lcccccccc}

\hline \hline
 &\multicolumn{3}{c}{H$_2$ lines (corrected from extinction)}& &\multicolumn{4}{c}{fit of PAH bands}  \\
\cline{2-4}
\cline{6-9}
           & $I_{S(2)} $     & $I_{S(3)} $      & $I_{S(3)}/I_{S(2)}$ & & $A_V$          & $I_{6.2}/I_{11.3}$& $\frac{[PAH^+]}{[PAH^0]}$ & $\frac{G_0\sqrt{T/10^3}}{n_H}$  \\
            &{\tiny  $(erg~cm^{-2} s^{-1} sr^{-1})$} & {\tiny$(erg~cm^{-2} s^{-1} sr^{-1}$)}&{\tiny $(K) $}        &                              & & &               &                     {\tiny (K$^{1/2}$ cm$^{-3}$)}     \\
\hline
 PDR1 & $ 2.3^{\pm 0.15} \times 10^{-4}$& $9.6^{\pm 0.1} \times 10^{-4} $   & $4.2^{\pm 0.30}$ & & 16  & 1.1 & 0.03 & 0.28  \\
 PDR2 & $1.6^{ \pm 0.1} \times 10^{-4}$  & $2.1^{\pm 0.1} \times 10^{-4} $  & $1.3^{\pm 0.15}$ & & 12   & 1.8 &1.60 & 0.49\\
 PDR3 & $1.8^{ \pm 0.1} \times 10^{-4}$  & $2.0^{\pm 0.1} \times 10^{-4} $  & $1.1^{\pm 0.15}$ & & 19  &  2.0 & 2.44 & 0.56\\
\hline
Mon R2$^{\dagger}$ & $8.0^{ \pm 0.1} \times 10^{-5}$ & $10.3^{\pm 0.1} \times 10^{-5} $ & $1.3^{\pm 0.10}$ & & 9 & 1.9 & 1.9 & 0.52\\
\hline
\multicolumn{8}{l}{ $^{\dagger}$spectrum obtained integrating the IRS cube over area imaged}
\end{tabular}
\end{center}
\end{table*}

\subsubsection{Estimating $G_0$ using PAH ionization ratio\label{gon}}

As shown by models \citep{tie05} and observations (e.g. \citealt{bre05, fla06, gal08}), 
the PAH ionization ratio ($[PAH^+]/[PAH^0]$) depends on the parameter $G_0 \sqrt{T}/n_e$ where $T$ is the gas temperature
and $n_e$ the electron density.
Ionization of PAHs will influence the emission cross section of these molecules, so that the ratio
between the 6.2 to 11.3 $\mu$m bands ($I_{6.2}/I_{11.3}$) will increase with increasing $G_0 \sqrt{T}/n_e$.
 Assuming that most electrons are provided by the ionization of carbon,
we can write that $n_e \simeq x(C^+) n_H \simeq [C]/[H] n_H$, with $[C]/[H]$ the elemental abundance of carbon
assumed to be $1.6 \times 10^{-4}$ \citep{sof04} . Finally, 
using the empirical law of \citet{gal08}, we can relate the $I_{6.2}/I_{11.3}$ to  $G_0(T/10^3)^{1/2}/n_H$ by:
\begin{equation}
G_0(T/10^3)^{1/2}/n_H \simeq  (1990~[C]/[H]) \times ((I_{6.2}/I_{11.3})-0.26)
\label{galiano}
\end{equation}
Using the ratio between the 6.2 and 11.3 $\mu$m bands for neutral
and ionized PAHs found in Table 1 in \citet{rap05} we  relate $I_{6.2}/I_{11.3}$ to the ionized to neutral PAH density 
ratio $\frac{[PAH^+]}{[PAH]^0}$, using Eq. (2) in \citet{job96}:
\begin{equation}
I_{6.2}/I_{11.3}=1.12 \times \frac{1+\frac{[PAH^+]}{[PAH]^0}\times0.85}{1+\frac{[PAH^+]}{[PAH]^0}\times0.29}.
\label{joblin}
\end{equation}

\section{Application to  Mon R2 PDRs}
The high value of I$_{S(3)}$ we observe require both $G_0 > 1\times 10^4$ and $n_H > 1\times 10^4$ cm$^{-3}$
(Fig. 2 and \citealt{bur92, kau06}). This implies that the observed $I_{S(3)}/I_{S(2)}$ ratio, after correction for extinction
(Table \ref{table2}), allows to derive the approximate densities of PDR 1,2,3 in Fig.\ref{pdr}-b.
In addition, since these lines are thermalized, the gas temperature, $T$, is
coincident with the rotational temperature, $T_{rot}$, and we can infer $T$ for the
different PDRs (Table \ref{table3}).  To calculate the rotational temperature we have
assumed an ortho-to-para ratio of 3, that is
the equilibrium value for temperatures larger than 100 K. Finally,
using the value of $G_0(T/10^3)^{1/2}/n_H$ estimated with the PAH ionization
fraction (or $I_{6.2}/I_{11.3}$), and the above derived $T$ and $n_H$ we can derive the intensity
of the radiation field that illuminates the three PDRs and position precisely PDRs 1,2 and
3 in Fig. 2. The found values for $n_H, T, G_0,$ (Table \ref{table3}) are consistent with Mon R2 
being a dense and highly UV irradiated PDRs as estimated from other molecular lines \citep{cho00,riz03, riz05}. 
The found values for density are lower than in these previous works,
because the present tracers (PAHs and H$_2$) probe the outermost cloud regions
directly exposed to the UV radiation field that are usually warmer ($>$500 K) and
less dense than regions situated deeper into the molecular cloud (see cut in Fig. 1).
Finally, the estimated parameters using this technique, for the \emph{entire} 
Mon R2 region (see Table 2), are more consistent with the ones derived for 
spatially resolved PDRs 2 and 3.


\begin{table}[ht!]
\caption{Derived physical parameters for PDR1, PDR2 and PDR3. Errors are the result of propagation from those in Table 1.}
\label{table3}
\begin{center}
\begin{tabular}{lccc}

\hline \hline
           & $n_H$ (cm$^{-3}$) & $T_{rot}$(K)  & $G_0$  \\
\hline
PDR1& $4.3^{\pm 0.5}\times 10^5$  & 574$^{+25}_{-22}$  & $1.6^{\pm 0.2 } \times10^5$ \\
PDR2& $4.0^{\pm 0.4}\times 10^4$  & 331$^{+19}_{-17}$ & $3.3^{\pm 0.3} \times 10^4$ \\
PDR3& $3.7^{\pm 0.3}\times 10^4$  & 314 $^{+18}_{-16}$ & $3.7 ^{\pm 0.2}\times 10^4$ \\
\hline
Mon R2 & $4.0^{\pm 0.4}\times 10^4$  & 321$^{+18}_{-16}$   & $3.7 ^{\pm 0.4}\times 10^4$  \\
\hline
\end{tabular}
\end{center}
\end{table}

\section{Conclusions}

The main results we have presented in this Letter are:
(\emph{i}) Mon R2 is a unique template of high-UV/high-density 
PDRs that is spatially resolved, similar to the Orion bar (where the UV field is slightly lower),
(\emph{ii}) our observations are consistent with PDR models which predict that
 in high-UV and dense PDRs, the excitation of pure rotational H$_2$ lines is 
collisional and depends on gas temperature,
(\emph{iii}) for this reason, the spatial distributions of PAHs and warm H$_2$
are different, contrary to what is seen in cool PDRs,
(\emph{iv}) however, because PAH emission 
is present in the extended region illuminated by the radiation,
their ionization fraction can be used to probe the
intensity of the UV radiation field, even where H$_2$ emission peaks.
Thus, the mid-IR spectrum of dense and highly irradiated PDRs
appears as an efficient probe of the physical conditions in these 
environments. The ratio between the molecular H$_2$ 0-0 S(3)/S(2) line intensities
allows to directly compute the density and temperature of the gas, 
while the measurement of the ionization fraction of PAHs allows to 
derive the intensity of the radiation field. 
We show that the derived parameters for the entire Mon R2 region
are consistent with those found for the spatially resolved PDRs 2 and 3.
This is likely because the dense PDR1 occupies only a small part of the whole 
region, and therefore dilution effects imply that the spatially averaged spectrum 
is dominated by emission from lower density PDRs.
Thus, we suggest that this spectral methodology could be useful to derive the dominant 
physical conditions in PDRs that are not spatially resolved in the mid-IR.
In particular, the PDRs at the surface of protoplanetary disks around
Herbig Ae$/$Be stars \citep{ber09} or in the inner rim of T-Tauri disks \citep{agu08}
are expected to be the targets where such an analysis can be applied. The mid-IR
spectrum of starburst galaxies is probably dominated by the emission of
PDRs having similar conditions as those described in this paper (see \citealt{fue08} for the case of M82). 
Thus, the present methodology could be a useful tool to derive global
properties of galaxies, in connection with massive
star formation activity, using forthcoming James Webb  and SPICA space telescopes 
that will observe these emission features at low and high redshifts.

\acknowledgments 

We thank the anonymous referee for his constructive criticism and careful reading of the manuscript.
This work is based on observations made with the Spitzer Space Telescope, which is
operated by the Jet Propulsion Laboratory, California Institute of
Technology under NASA contract 1407. OB is supported by JAE-Doc CSIC fellowship.
JRG was supported by a Ramon y Cajal research contract from the spanish
MICINN and co-financed by the European Social Fund.
OB, PP and CJ acknowledge the french national program PCMI.

\bibliographystyle{apj}

\end{document}